\documentclass[preprint,12pt]{elsarticle}

\usepackage{amssymb,amsmath,url}

\newcommand{\mn}{\mathbb{N}} 

\newcommand{\var}[1]{\text{var}\left(#1\right)}

\newcommand{\Esym}{\text{E}}

\newcommand{\E}[1]{\Esym\left[#1\right]}

\newcommand{\Probsym}{\mathbb{P}}

\newcommand{\Prob}[1]{\Probsym\left[#1\right]}

\newtheorem{defn}{Definition}

\journal{Computer Communications}
\begin{document}
\begin{frontmatter}

\title{A critical look at power law modelling of the Internet}

\author{Richard G. Clegg}
\address{Department of Electronics and Electrical Engineering,
University College London, \\email: richard@richardclegg.org }

\author{Carla Di Cairano-Gilfedder}
\address{BT Research, \\email: carla.dicairano-gilfedder@bt.com}

\author{Shi Zhou}
\address{Department of Computer Science,
University College London,
\\email: s.zhou@cs.ucl.ac.uk}

\begin{abstract}
This paper takes a critical look at the usefulness of power law models
of the Internet.  The twin focuses of the paper are Internet traffic
and topology generation.  The aim of the paper is twofold.
Firstly it summarises the state of the art in power law modelling
particularly giving attention
to existing open research questions.  Secondly it provides insight into
the failings of such models and where progress needs to be made
for power law research to feed through to actual improvements in
network performance.
\end{abstract}




\end{frontmatter}


\section{Introduction}

Power laws describe a wide range of phenomena in nature and a large body of
ongoing research investigates their applicability in fields such as computer
science, physics, biology, social sciences and economics. Power
law distributions are characterised by a slower than exponentially decaying
probability tail, which loosely means that large values can occur with a
non-negligible probability (see the next section for formal definitions).
They can be used to
characterise a variety of relations such as for example the distribution
of income, city
population, citations of scientific papers, word frequencies,
computer file sizes and the number of daily hits to a given website.  See
\cite{mitzenmacher04} and \cite{newton05} and references therein for further
examples.

The aim of this paper is not to be a general survey of power laws in networks
but instead to be a critical look at open questions and the outcome of such
research, in particular with regard to the question ``How can power law
modelling improve network performance?"  The paper looks at two
separate areas where power law research has been of interest in
the Internet.
The study of power laws in the analysis of Internet
traffic characteristics
has been ongoing since 1993 and in
Internet topology generation since 1999.

In 1993, the seminal paper \cite{leland1993}
(expanded in \cite{ltww94}) found evidence of the existence of power law
relationships in network traffic by observing long-range correlation in Local
Area Network (LAN) traffic. This brought the concept of self-similarity,
and the related concept of Long-Range Dependence (LRD), into the field of
network traffic and performance analysis. Before this finding, network
traffic and performance studies had been mainly based on models, such as
Poisson processes, which assume that traffic exhibits no long-term correlation.
In networks with long-range correlated traffic, queuing performance can
very different to that of traffic assumed independent or only having short-term
correlations.
Subsequently power law relationships have been observed in several other
contexts on many different types of network.

In 1999 it was also discovered that the global Internet structure is characterised by a power
law~\cite{Faloutsos99}.  That is, the probability distribution of a node's connectivity (measured for example
by the number of BGP peering relations that an autonomous system has) follows a power law. This discovery
invalidated previous Internet models that were based on the classical random graphs. Since then a lot of
efforts have been put into studying the Internet power law structure
\cite{strogatz01,krapivsky01,albert02,bornholdt02,subramanian02,Chen02,dorogovtsev03a,caldarelli03,Pastor04,chang06}.

This paper reviews the measurements and models of the Internet topology, and
comments upon whether the power law is in itself an adequate characterisation
of the system.  It questions whether models based on power laws provide a suitable
platform for theoretical and simulation analysis of the Internet's traffic and
topological characteristics.  Finally, it provides discussion of how such
research could be of use in improving network performance (which, after all,
should be the ultimate goal of networking research).

The structure of the paper is as follows.
Section \ref{sec:defns} provides
the basic mathematical definitions used throughout the paper:
heavy-tailed distributions, long-range dependence and statistical
self-similarity.
Section \ref{sec:traffic}
describes the use of power
law relationships to model the statistical nature of Internet traffic.
Section \ref{sec:topology} discusses ``scale-free networks" a
power law relationship which describes the connectivity of networks.

\subsection{Basic definitions} \label{sec:defns}

In the sense meant in this paper, a power law relationship is a
function,
$f(x)$ with the form $f(x) = \alpha x^ \beta$ where $\alpha$ and
$\beta$ are non zero constants.  Several relationships
of interest in the Internet have been shown to have this form
asymptotically
(usually as $x \rightarrow \infty$).

\begin{defn}
A random variable $X$ (which may be continuous or discrete)
is said to have a {\em heavy-tailed\/} distribution
if it satisfies
\begin{equation*}
\Prob{X > x} e^{\varepsilon x} \rightarrow \infty, \text{ as }
x \rightarrow \infty.
\end{equation*}
\end{defn}

Often a specific power law form is assumed for the distribution:
$$
\Prob{X > x} \sim C x^{- \alpha},
$$
for some $C > 0$ and some $\alpha \in (0,2)$.  The symbol $\sim$ here
and for the rest of this paper means {\em asymptotically equal to\/},
that is
$f(x) \sim \phi(x) \leftrightarrow f(x)/\phi(x) \rightarrow 1$ as $x
\rightarrow
\infty$ (or occasionally, some other limit).

\begin{defn}
Let $\{X_1, X_2, \dots \}$ be a time series.  The
series is {\em weakly-stationary\/} if it has a constant and
finite mean ($\E{X_i} = \mu$ for all
$i$, where $\Esym$ means expectation) and for all $i, j \in \mn$
the covariance between $X_i$ and $X_j$
(i.e. $\E{(X_i - \mu)(X_j - \mu)}$) depends only on $|j - i|$.
\end{defn}

Weak stationarity is assumed for much that follows but in practice
is not met by real network traffic over all timescales (for example,
over a sufficiently long time the mean traffic level is not
stationary, it varies with daily and weekly
periodicity)
and may not be met at all \cite{cleveland2000,cao2001}.

If the time series is weakly-stationary then the Auto Correlation
Function (ACF) $\rho(k)$ is given by
\begin{equation*}
\rho(k) = \frac{\E{(X_t - \mu)(X_{t+k} - \mu)}}{\sigma^2},
\end{equation*}
where $\mu$ is the mean and $\sigma^2$ is the variance.

The ACF allows the definition of
{\em long-range dependence\/} which is sometimes called
{\em long memory\/} or {\em strong dependence\/}.
A standard reference on the topic is \cite{b94}.
A commonly used definition is the following.
\begin{defn} \label{defn:lrd}
A weakly stationary time series is {\em long-range dependent\/} if the sum
of the autocorrelation over all lags $\sum_{k=1}^\infty \rho(k)$
diverges.
\end{defn}


\begin{defn} \label{defn:hurst}
The {\em Hurst parameter\/} is a commonly used measure
of LRD.  This makes the assumption that the ACF follows
the specific functional form
\begin{equation}
\rho(k) \sim C_\rho k^{-\alpha} = C_\rho k^{2-2H},
\label{eqn:lrdacf}
\end{equation}
where $C_\rho > 0$ and $\alpha \in (0,1)$ and
$H \in (1/2,1)$ is the {\em Hurst parameter\/}.
\end{defn}

Note that sometimes it is this and not Definition \ref{defn:lrd}
which is taken as the definition of LRD.  Other measures of LRD
include Hurstiness \cite[Chapter 8]{ganesh2004} and the
`strength' parameter used by \cite{cao2001}.

LRD processes
which meet Definition \ref{defn:lrd} but not Definition \ref{defn:hurst}
will have no well-defined Hurst parameter.
The value $H=1/2$ is usually taken to mean
independent or short-range dependent data.
Values of $H \in (0, 1/2)$ are sometimes
termed anti-long-range dependence.
Values of $H \leq 0$ or $H \geq 1$ do not give useful models
\cite[Section 2.3]{b94}.

LRD can also be considered in the frequency domain.
In this case, the characteristic of LRD is a pole in
the spectral density (usually at zero).



\begin{defn}
Let $Y_t$ be a stochastic process in continuous time $t \geq 0$.
The process is {\em exactly self-similar\/} with self-similarity
parameter $H$
if for any choice of
constant $c > 0$, the rescaled process $c^{-H}Y_{ct}$ is
equal in distribution to the original process $Y_t$.
\end{defn}
Note that a similar definition can be given for discrete time
stochastic
process.

\begin{defn} \label{defn:2nd_order_ss}
Let $Y_t$ be a stochastic process and $Y^{(m)}_t$ be the process
derived from it by $Y^{(m)}_t = \frac{1}{m} \sum_{i=tm - (m-1)}^{tm}
Y_i$.
A process $Y_t$ is {\em exactly second-order self-similar\/} if,
for all $m$, the process $\{m^{1-H}Y^{(m)}_t\}$ has the same
variance and autocorrelation function as $Y_t$.  That is to say, for
all $k \in \mn$ and $m \in \mn$,
\begin{equation*}
\var{Y_k^{(m)}} = \frac{\var{Y_k}}{m^{2-2H}}
\end{equation*}
and
\begin{equation} \label{eqn:2nd_order_ss}
\rho^m(k) = \rho(k),
\end{equation}
where $\rho(.)$ is the ACF of $Y_t$ and
$\rho^m(.)$ is the ACF of $Y^{(m)}_t$.
\end{defn}

Second-order self-similarity can also be defined in terms of the
second central difference operator \cite{kriesten1999}.
A process $Y_t$ is {\em asymptotically
second-order self-similar\/} if
\eqref{eqn:2nd_order_ss} holds as $k \rightarrow \infty$.

Finally it remains to define scale-free (or power law) networks.
\begin{defn} \label{defn:scale_free}
Let $G$ be an undirected graph.
Let $P_k$ be the probability that a randomly selected node in
$G$ has degree $k$.  The graph $G$ is {\em scale-free\/} if $P_k$ (the
node-degree distribution) is heavy-tailed if:
\begin{equation*}
P_k \sim C k^{-\alpha},
\end{equation*}
where $C >0$ is a constant and $\alpha \in (0,2)$.
\end{defn}

Similar definitions can be constructed for a directed graph.  The
in-node
degree distribution and the out-node degree distributions are treated
separately in this case.

A process which scales in a constant way is sometimes referred to
as {\em mono-fractal\/}. A generalisation of this
is a {\em multi-fractal\/} process which exhibits complex behaviour
that changes over different timescales \cite{pw00}.  When the
multi-fractal behaviour can be approximated by a combination of
two (or a small number of) monofractals then the process is sometimes
described as having monofractal behaviour at different timescales rather
than multifractal behaviour.

There are many connections to be made between these power laws,
some more obvious than others.  For example, a scale-free network
is simply an example of a heavy-tailed distribution (as its node-degree
distribution is heavy-tailed).

One connection which is
sometimes less than clear from the
literature is that exact second-order self-similarity
as in Definition \ref{defn:2nd_order_ss} implies
LRD of the form given by \eqref{eqn:lrdacf}.  LRD of
the form in \eqref{eqn:lrdacf} implies asymptotic self-similarity.
The details of this relationship can be found in \cite{kriesten1999}
and \cite{gubner2005}.
There is a more subtle connection between self-similarity and
long-range dependence.
If a self-similar process $Y_t$ has stationary increments
and $H \in (0,1)$ then it can be shown (see \cite[page 51]{b94})
that the increment process given by $X_i = Y_i - Y_{i - 1}$ for $i \in
\mn$
has an ACF given by
$\rho(k) \sim H(2H-1)k^{2H - 2}$,
which implies that for $H \in (1/2,1)$ then the increment process
is long-range dependent.

The connection between heavy-tails and long-range dependence
is more subtle.
One such connection is \cite[Theorem 4.3]{heath1998} which states that
%
in an on/off process with heavy-tailed on periods
and off periods which fall off faster is a long-range dependent process.
%
Other connections between power laws can be found in
\cite{cb95,wtsw97}. These papers show that multiplexing a high number of
independent on/off sources with heavy-tailed strictly alternating on
and/or off periods gives rise to self-similarity.

\section{ Power laws and Internet traffic} \label{sec:traffic}

Nearly fifteen years ago, the seminal paper \cite{leland1993}
found the existence of power law behaviour in Internet
traffic. A time series describing LAN
Ethernet packet traces at Bellcore showed evidence of second order self-similarity or
long-range dependence.  This paper for the first time questioned
traditional modelling assumptions
and showed that existing models (often based on Poisson processes)
would not correctly estimate important characteristics of a network.
Since this paper, many hundreds of papers have been published about
the power law behaviour of Internet traffic.  A recent
edition of the journal {\em Performance Evaluation\/} was devoted to
this topic and the editorial describes modelling of LRD and heavy-tails
as ``One of the most important research topics in performance modelling
and evaluation in the last decade" \cite{liu2005}.

\subsection{ Measuring long-range dependence}

The Hurst parameter is often used as an estimate of traffic's LRD. This parameter however
has to be
used with prudence, as measuring traffic LRD and statistical self-similarity is
a complex task which may be affected by many factors. Although, the estimation process can
provide indication of the existence of long-range dependent characteristics, it does not
unequivocally prove the existence of authentic LRD, as these characteristics may simply be
due to traffic non-stationarity. In the time domain, the estimation of the Hurst parameter
is characterised by the fall off
of the ACF at high lag.  However, the high lag measurements are those
at which the fewest readings are available and the data is most unreliable.  Similarly,
in the frequency domain, the LRD is characterised by the behaviour
of the spectrum at frequencies near zero, which are necessarily hard frequencies
to measure.
In terms of queuing performance, despite the common misconception, a high
Hurst parameter does not always lead to worse performance
or longer queues \cite{neidhardt98concept}. In fact, depending on the timescales of
interest, traffic with a high Hurst parameter can lead to better performance than traffic
with a low Hurst parameter.
No single Hurst parameter estimator can be considered infallible, as this can
hide LRD when it exists or create it when it does not \cite{kara04tenyears}. In addition, the Hurst parameter
itself expresses the traffic scaling of the fluctuations around the mean and does not
measure traffic burstiness.

It is certain that simply examining the ACF is not a robust way to estimate the
Hurst parameter.   In addition, a number of biases may be present in real-life data
which could cause problems. These include periodicity (users and processes daily
usage patterns) and trends (traffic volume changes throughout the measurement
period) which violate the assumption of weak-stationarity. The topic of measuring
LRD is beyond the scope of this paper, the reader is referred to
\cite{taqqu1997,bardet2003} for work which compares existing techniques.


\subsection{Evidence for and against power law behaviour in Internet
traffic}

The original long-range dependence findings reported in 1993 \cite{leland1993} have
subsequently been replicated in many different studies.
In 1995, Floyd and Paxson \cite{pf95} found that WAN traffic is also
consistent with self-similar scaling.
These findings have been confirmed in the late nineties in \cite{cb97,fgw98}. In
particular, in \cite{cb97} the authors analyse WWW traffic
and observe self-similarity in the patterns of recorded traffic
and a heavy-tailed distribution in the sizes of the files
transferred.  They claim that heavy-tailed sizes of transferred files
is the cause of the observed self-similarity.
Also, in the late nineties, evidence was found that heavy-tailed
distributions characterised a number of different measurements
related to network traffic.  In \cite{pkc96} the
authors report on observations of heavy-tailed distributions of
file sizes on web servers and also of CPU time taken by processes.

The paper \cite{hernandez04} analyses WWW flow duration
distribution at a lightly utilised academic campus Internet access.
It finds that the tail of the flow duration distribution
does not stabilise.
The suggestion is that the best fit to the data is with a power law
which varies in time.

In 2005, \cite{xia2005} also investigated the power law behaviour of WWW traffic
and found evidence of self-similarity over a number of timescales.


%
%

The paper \cite{cao2001} is sometimes cited as evidence that LRD is not an important
property of Internet traffic.
The data they analyse was collected in 2000 on a 100 Mbps Bell Labs Ethernet link.
%
%
Looking at inter-arrival times the authors find that when the traffic
has more connections present the ``strength" of the LRD is decreased.
Note that this ``strength" is not related to the Hurst parameter but
could be considered analogous to the proportion of the traffic which
exhibits LRD.  Their
conclusion is that as the number of connections in the network increases the traffic
will remain long-range dependent, but that the strength of the LRD will be weaker, and the
arriving traffic will look more like a Poisson process.

In 2003, observations were recorded
on university access links \cite{park05changetraf}.
In the majority of the traces, the packet and
byte count time series exhibit intermediate to heavy LRD, regardless of
time of day or day of week. LRD is also found to be unaffected by
traffic load and number of active connections. Therefore, in these
access links, multiplexing of an increasing large number of TCP
flows did not reduce correlation.

\subsection{Behaviour at different timescales}

Some authors have claimed that different scaling behaviour occurs
at different time scales.  This matter still seems to be an open research question.
LRD and self similarity are both
``monofractal" models in the sense that they assume a constant scaling behaviour
over all time-scales.  Strictly speaking asymptotic self similarity and LRD only
imply this behaviour in the limit (at high lags or low frequencies). 
Multi-fractal modelling allows
this scaling behaviour to change at each time scale considered.
The topic of multi-fractal modelling is beyond the scope of this paper.
For a good introduction see \cite{riedi2003}.  Some authors have
claimed that a multi-fractal
approach is necessary to replicate the behaviour of Internet traffic.  However,
others have argued that this is not the case and a mixture of different monofractal
scalings at different timescales is necessary.

It has been argued (see \cite{rl97,fgw98})
that protocol mechanisms (such as the TCP feedback mechanism)
have the greatest impact at smaller timescales. At these
timescales they claim that the traffic is consistent with multi-fractal scaling,
but at larger timescales (larger than the typical RTT on the network being investigated)
the traffic looks self-similar.

\cite{vehel97} compares the scaling behaviour of aggregated fractional
Brownian motion processes
with that of real traffic and concludes that it is not a good match
and therefore suggests multifractal behaviour may be necessary to provide
a good fit to real traffic traces.

In \cite{erramilli00performance} the relationship between wide-area
traffic correlation and link utilisation is explored at different
timescales. They find that at small timescales burstiness can impact
on performance at low and intermediate utilisations, while
correlations at larger timescales are more significant at intermediate
and high utilisations.

The paper \cite{zhang03smalltime} analyses backbone traffic traces
at multiple Tier 1 links and investigates its behaviour at
small scales (less than one second).
The presence of correlation at small timescales is attributed to the
characteristics (and not the number) of the aggregated flows and
affected by the presence of dense flows characterised by bursts of clustered
packets. They conclude that, at small timescales, traffic has
mainly a monofractal behaviour and even that the traffic is
``almost independent" -- this is a
contrast to much of the other work discussed in this section.

The paper \cite{jiang04timeorigin} sheds more light as regards
to the possible causes of traffic correlation at sub-RTT timescales on
backbone links. This paper confirms the findings in
\cite{zhang03smalltime}
but in addition suggests that these clusters of bursts
derive from TCP self-clocking mechanism and queuing delays.
These cluster of bursts are produced by flows with large bandwidth-delay
product relative to their window size.

Internet backbone traffic dated 2002--4 is also analysed in
\cite{kara04tenyears} with the
conclusion that the Poisson distribution can adequately
model packet arrivals at smaller timescales (the threshold where behaviour
changes from Poisson to LRD varies but is around 1000ms).
It confirms the existence of LRD in packet and byte
counts at timescales larger than a second.

In \cite{uhlig2004}, the author considers the rate of TCP flow arrivals
rather than the total traffic on a link.  Several traces are investigated,
collected between 1993 and 2002.  The analysis finds different scaling
behaviours over a range of timescales and concludes that the
flow arrivals are uncorrelated at the smallest timescales,
correlated at timescales between seconds and minutes and consistent
with
``LRD or self-similarity between minutes and hours" but non-stationary
time-of-day behaviour prevails at longer time scales.

In \cite{hohn2005} the authors argue using analysis of several traces (taken between
1989 and 2002)
that at longer timescales LRD is an appropriate model and at shorter
timescales they refer to
the behaviour as ``pseudo-scaling", a process which gives the appearance of
multifractality but ``which does not have true multifractal scaling underlying it" --
in other words that the multifractal scaling observed by earlier authors is
unnecessary.
The scale at which the behaviour changes differs according to the trace being
examined.

In summary, consensus seems to have formed that LRD
behaviour predominates when traffic is considered at a larger timescales (at least
until the user related non-stationarity disturbs the observation).
However, the shorter timescale behaviour is a matter for much debate with some authors suggesting something as simple as a
Poisson model is adequate, others
suggesting multi-fractal models are necessary and many taking positions in between these
extremes.

\subsection{Possible causes of long-range dependence in network traffic}
\label{sec:causes}

The origins of power law behaviour in network traffic have not been unequivocally
identified. Several possible causes for the presence of long-range dependence have been
proposed in the literature.

\textit{\textbf{Heavy-tailed distributions}}.  A common suggestion is that the
heavy-tailed nature of data transfers leads to LRD in the resultant traffic
(see \cite[Theorem 4.3]{heath1998})
%
Simulation studies have confirmed this experimentally: \cite{pkc96}, for example, found it
to hold over a range of link bandwidths and a range of buffer sizes. Also in \cite{fghw99}
simulations show that heavy-tailed file sizes lead to self-similarity on large timescales
but also that the delay behaviour interacts with the TCP feedback
mechanism to greatly alter the structure of the traffic at shorter
time scales.

\textit{\textbf{TCP protocol}}.
It has been argued \cite{vb00} that TCP congestion control alone can cause self-similarity
regardless of the application layer traffic characteristics. This argument however is
contested in
\cite{figuerdo2005} which looks at the same data but over longer time scales and finds
that it is not consistent with power law behaviour. Also \cite{hohn2002}, by shuffling
network samples, reordering traffic, and removing the effects of TCP mechanisms while
leaving the effects of heavy-tailed traffic, is able to show that it is heavy-tailed
traffic rather than TCP feedback mechanisms which leads to long-range dependence.

It has been suggested \cite{figueiredo2000auto} using evidence
based upon Markov modelling that the TCP timeout mechanism can lead to
``local long-range dependence" which they also refer to as ``pseudo self-similarity",
that is to say, self-similarity over a small number of timescales (note that this is not
true self-similarity).

The proposal in \cite{pe97}
suggests that TCP retransmission mechanism can give rise to self-similarity.
Also \cite{pkc96}
concludes that TCP can preserve long-range dependence over time while
\cite{veres03tcps} suggests that TCP can preserve correlation over space.

\textit{\textbf{Queuing/Routing effects}}.
Another possibility is that power law traffic arises as a result of the interaction of
queues and routing on a network \cite{borella98measurement}. Simulation experiments
shown that even when ``packet inter-departure times are independent,
arrival times at the destination exhibit LRD'', perhaps as a result of the routing
algorithms \cite{S&V,arrowsmith2002}.
%

\textit{\textbf{Multi-layers and timescales}}. There is also the possibility that
long-range dependence simply arises because of the combination of processes occurring
at different timescales: user's activity, session, and transmission processes. In
fact, it can be seen that even under the assumption of Poisson distribution for  all
usage, session, and transmission processes, the mere presence of multiple layers may
lead to correlated traffic \cite{rm99}.

\textit{\textbf{ Intrinsic traffic nature}}. It has long been known that some types of
traffic exhibit LRD at the source.  For example, variable bit rate video traffic deriving
from a single flow shows LRD in a time series of traffic \cite{gw94}.

While no clear consensus has yet formed, many of the authors cited in this and
the previous two
sections agree that heavy-tails are the cause of the LRD observed in larger time
scales.  No consensus seems yet to have been reached on the behaviour of traffic
at shorter time scales and this remains an important topic for traffic research.
The lack of consensus in this is reflected in the number of possible causal
models for the short timescale behaviour.

\subsection{Effects on queuing}
\label{sec:queuingeffects}

The effects of long-range correlated traffic on buffer dimensioning
have been analysed by means of appropriate queuing models developed in
\cite{azn95,n94,pm96,tg98,kherania2005} among others. These models apply to infinite
buffers and only provide asymptotic results.
Under the assumption of infinite length buffer and long-range dependent input traffic the main
finding is that the distribution of queue length has slower than
exponential decaying tail, as opposed to exponential observed for short-range dependent traffic.
This decaying function has instead been
described by other distributions such as for example a Weibull \cite{n94} and polynomial
\cite{tg98}.

In the case of finite buffer systems, it has been suggested that, in a network with
long-range dependent traffic, the packet loss ratio is several orders of magnitude higher
than with short-range dependent traffic \cite{chen95model}. The packet loss ratio could
only be contained by choosing very large buffers which would have an impact on queuing
delay \cite{chen95model}.
%
%
%
However, in the mid-nineties other authors cast doubt on the usefulness of
power law models of Internet traffic, by questioning the importance of capturing traffic
long-range dependence in the case of finite buffers \cite{gb96,re96}. They argue that
correlation becomes irrelevant for small buffers and short timescales.



\subsection{ Traffic generation models }
\label{sec:lrdgenmodels}

A variety of mathematical models have been suggested in the literature to capture the LRD
in Internet traffic.  For a comprehensive review of these models the
reader is referred to \cite{pw00}.  Only a short summary is provided here.

{\em Fractional Brownian motion\/} (fBm) is a non-stationary
stochastic process which is a
generalisation of the well-known Brownian motion,
but with a dependence term between samples.  It is a self-similar
process and has a defined Hurst parameter H, with the Brownian motion obtained for $H =
1/2$.
If $B_H(t)$ denotes the fBm then the difference process $Y_k(\cdot)$ defined as
$Y_k(t) = B_H(t+k) - B_H(t)$ with $H \in (1/2, 1)$ is the
{\em fractional Gaussian noise\/}(fGn) which is long-range dependent.
Several methods exist for generating a fGn process, for example
\cite{paxson1997}.

Although fGn is mathematically attractive its simplicity
means that it cannot capture a diversity of mathematical properties.
The queue length distribution obtained with a fGn process decays according to the Weibull
or ``stretched exponential'' distribution, which is heavy-tailed only in a weak sense \cite{roberts96}.

{\em Fractional Auto-Regressive Integrated Moving Average\/}
(FARIMA) \cite[pages 59--66]{b94}
models are an expansion of the classic time-series ARIMA models and allow
modelling of long and short range dependence simultaneously and independently.

Long-range dependence can also be generated by using {\em chaotic maps} as first
proposed by Erramilli and Singh \cite{erramilli94chaotic}.
%
%
However, modelling based on chaotic maps requires considerable experimentation, as these are
very sensitive to initial conditions and their many parameters' estimation is often a
complex task \cite{rm99}.
The queue length distribution obtained with a chaotic maps family has been found to
decay according to the Weibull distribution \cite{pruthi95}.

Another model is based on the superposition of {\em heavy-tailed on/off\/} sources
\cite{wtsw97}.  The process obtained by multiplexing many on/off sources with heavy-tailed
distributions tends to a fGn process.

Finally, another technique for modelling traffic is by means of {\em Wavelet\/} analysis
\cite{burrus98}.
This allows not only capturing the Hurst parameter but
also synthesising a wide range of scaling behaviours and the
replication of the multi-fractal spectrum \cite{riedi1999,riedi2003}.

An important criticism of these models is in their replication of queuing
behaviour.  While much work has been done to show that the models replicate
certain representative traffic statistics, one of the primary motivations
cited for using LRD models of queuing is estimating delays and buffer
overflow probabilities.  These models have not been shown to do this well,
indeed while it has been shown that some mathematical models of LRD have
very different queuing behaviour to non LRD versions of those models,
it remains to be shown that LRD is necessary to replicate the queuing and
delay performance of real traffic.



\subsection{Criticisms and commentary}

%

Although the majority of papers appear to replicate the finding that LRD is present
in network traffic, some have questioned whether other models are more appropriate
(for example multi-fractal models \cite{rl97,fgw98}). Multifractals are in fact able
to model varying scaling behaviour over different timescales,  as they are
characterised by a time dependent scaling coefficient. In addition, LRD appears at
long timescales which are more relevant for network dimensioning and less for
queuing behaviour. Others have also questioned whether LRD may be unimportant in
practice, for example due to multiplexing gains \cite{cao2001}.

Consensus seems to be forming on the origin of LRD behaviour (as discussed
in section \ref{sec:causes}) although some controversies remain.
As regards to its effects, papers in the area often focus on the fact
that LRD may impact on network performance by increasing delays or
increasing the packet loss expected for a given buffer size.
However this relationship is not a simple one and the presence of
LRD does not always have a negative impact \cite{neidhardt98concept}.
If a cause were unequivocally established the question would
remain, ``how might we go about eliminating LRD from the network
given this cause?"
If heavy-tailed
file transfers are the cause then no clear method for resolving the
problem is obvious.
However, if TCP feedback mechanisms are a cause it would be difficult
to change this without changing the protocol itself.

In order to understand the usefulness of power laws for practical studies,
an important question to ask is whether LRD models generate traffic
with the same queuing properties as real Internet traffic.  If the
models from Section \ref{sec:lrdgenmodels} are to be useful then,
when correctly tuned to the parameters of a genuine packet trace,
they should have the same mean delay and buffer overflow probabilities
as the genuine traffic.  Huebner et al \cite{huebner1998}
tested a Poisson model, a Weibull model, an autoregressive (AR(1)) model, a
Pareto model, and a Fractional Brownian Motion model for generating
traffic.  None of the models tested produced a good match for queuing
performance in all circumstances.  The fBm model was useful only when the
buffer size considered was large.
Similarly,
\cite{clegg2007} tests the queuing performance of fBm and three other
LRD models based on Markov modulated processes as well as some non LRD
models.  The models
are tuned so that their parameters (mean and Hurst parameter) match
real network traffic and the queuing performance of each is tested
in an infinite buffer simulation.  In this case, none of the traffic
models replicated the queuing performance of the real traffic and the
LRD models often showed different performance from each other despite
having the same mean and Hurst parameter.  Of course,
even if a model could be found which accurately reproduced
a given queuing  behaviour  obtained with  real traffic, this would
not solve the entire problem since the statistical nature of
Internet traffic arises at least in part from TCP feedback mechanisms,
which in turn depends upon potentially changing traffic levels and congestion.


Theoretically, some interesting queuing theory results for systems with
LRD input traffic have been achieved
but these results are often asymptotic results for infinite buffer models.
How applicable these would be in practical situations
remains an open question although, of course, it may be hoped that
future theoretical results will build on them.

Several questions therefore remain about LRD.  Which LRD model, if any,
is appropriate to generate traffic which has similar delay and buffer
overflow probabilities to real Internet traffic when queued?  Can
future networks be designed to mitigate the potentially deleterious effects
on performance which are said to result from LRD?  Can analytical models
be developed which give strong enough results to be practically applicable
to real traffic on real networks?

\section{Modelling Internet topology }
\label{sec:topology}

Topology is the connectivity graph of a network, upon which the network's physical and engineering properties
are based. The Internet contains millions of routers, which are grouped into 
tens of thousands of
sub-networks, called Autonomous Systems (AS). The Internet topology can be studied at the router level and
the AS level. Studies of the Internet topology very much depend on the availability and quality of measurement
data. In the last decade a number of projects have provided more and more complete and accurate data on
the Internet AS connectivity. By comparison it is more difficult to obtain router level data. So far
there are more studies on the AS-level Internet topology than on the router-level.

In this paper the Internet topology is considered only
at the AS level, in which a node is an AS network owned
by an entity with a large Internet presence, such as an ISP or a large company; and a link represents a peering relationship 
between two AS nodes in the border gateway protocol
(BGP)~\cite{quoitin03}. Research on the structure and evolution of the Internet AS graph is relevant because
the delivery of data traffic through the global Internet depends on the complex interactions between AS that
exchange routing information using the BGP protocol.

\subsection{Measuring Internet topology}

Measurements of the Internet AS graph have been available since the 
late 1990s. There have been two types of
measurements using different methodologies and data sources.

{\em Passive measurements\/} are constructed from BGP routing tables which contain information about links
from an AS to its immediate neighbours. 
The Routing Information Service of RIPE~\cite{ripe} is another
important source of BGP data. The widely used BGP AS graphs are produced by the National Laboratory for
Applied Network Research \cite{nlanr} and the RouteViews Project at 
the
University of Oregon \cite{oregon}. They
are connected  to a number of operational routers on the Internet for the purpose of collecting BGP tables.
The Topology Project at the
University of Michigan \cite{michigan} has provided an extended version
\cite{chang04} of the 
BGP AS graph by using additional data sources, such as the Internet Routing Registry (IRR)
data and the Looking Glass (LG) data. BGP-based AS measurements may contain links that do not actually exist
in the  Internet, but a more 
serious problem is that the BGP measurements may miss a significant number of
links~\cite{he07}.

{\em Active measurements\/} are based on the traceroute tool which sends 
probe packets to a
given destination and captures the sequence of IP hops along the forward path from the source to the
destination. The Internet research organisation CAIDA \cite{CAIDA} has developed a tool called {\em
skitter\/} which probes around one million IPv4 addresses from 25 monitors around the world. Using the core
BGP tables provided by RouteViews, CAIDA maps the IP addresses in the gathered traceroute data to AS numbers
\cite{murray01} and constructs AS graphs on a daily basis. DIMES \cite{DIMES} is a more recent large-scale
distributed measurement effort. It collects traceroute data by probing from more than $10,000$ software clients,
installed by volunteers in 
over $90$ countries, to destinations assigned by a central server at random from a set
of five million destination addresses. To further improve the completeness, DIMES merges the resulting AS
graph with that of RouteViews. By using more monitors and a larger list of distinct addresses, DIMES
produces larger AS graphs than skitter. The shortcoming of the traceroute measurements is that the
translation from IP addresses to AS numbers is not trivial and could introduce many errors~\cite{hyun03} and also, increasingly, firewalls block the probe
packets. A
recent study~\cite{Oliveira07} suggested that traceroute measurements should probe destinations more
frequently and avoid using a fixed list of destination addresses.

\subsection{Power law degree distribution}

In graph theory, degree $k$ is defined as the number of links, or immediate neighbours, of a node. Degree is
the principal parameter for characterising network connectivity. The first step in describing and
discriminating between different networks is to measure the degree distribution~$P(k)$, which is the
probability of finding a node with degree $k$. 
In 1999 it was discovered that the Internet topology
at the AS level (and the router level) exhibits a power law degree distribution $P(k)\sim C k^{-\gamma}$
\cite{Faloutsos99}, where $C > 0$ is a constant and the exponent $\gamma\simeq 2.2\pm0.1$. This means on the
Internet AS graph, a few nodes have very large numbers of links, whereas the vast majority of nodes have only
a few links. Although different Internet AS graphs produced from different data sources vary in the numbers
of nodes and links, all the Internet AS graphs are well characterised by a power law degree distribution
\cite{mahadevan05b}. The power law distribution is an evidence that the Internet AS level topology has
evolved into a complex, heterogeneous structure that is profoundly different from Internet models
based on the random graph theory. This discovery profoundly changed the understanding of Internet
topology. Since then there has been an international effort in characterising and modelling the Internet
topology.

\subsection{Power law or sampling bias?}

A major problem of current measurements of the Internet AS graph is that these measurements, whether
based on BGP, traceroute or other sources, miss a significant number of links \cite{chang04,cohen06,he07}.
Some researchers suggested~\cite{cohen06,he07} that there could be as many as $35\%$ of the links in the AS
level Internet that were still to be discovered. A series of papers~\cite{Lakhina03,Clauset05} reported that the
traceroute type of measurement data collected from a small number of observers are not only incomplete but
are possibly biased in such a way that graphs which in fact have Poisson degree distributions appear to
exhibit a power law. There has been a debate on whether the power law degree distribution an integral
property of the Internet AS graph or merely an artifact due to biased sampling methods.

There are two sides of the argument. Many researchers believe that the power law is an integral property of
the Internet. Firstly all Internet AS graph measurements exhibit a power law degree distribution including
the DIMES data which are collected from numerous observers distributed in thousands of AS networks around the
world, as well as the BGP AS graph based on routing table data collected from many monitors and accumulated
over many years. Secondly, a recent study \cite{DallAsta06} shows that if the larger real graph had a Poisson
degree distribution and the observed power law were due to sampling bias, then the real graph's average
degree would be very large. In the case of the AS network the true average
degree would have to be around one hundred.  The observed
average degree in the known sources is between five and seven so if this model were true it would require the
unlikely proposition that less than one in ten edges have been observed. Surely this can not be true.

On the other hand, there are also many researchers who are sceptical about 
the power
law degree distribution.
Firstly, the visibility of the 
AS graph can be influenced to a great extent by which vantage points are used, not
by how many. Secondly the analysis in \cite{DallAsta06} rejects the claim that the real AS graph may follow a
Poisson degree distribution, but the real question is whether the Internet AS graph is characterised by a
power law distribution or a different heavy-tail distribution which does
not follow a power law.

Only better measurement data can settle this issue. The current situation is that all measurements
are incomplete and bias in one way or another. There is an urgent need for improved methods to produce more
complete and accurate data. A recent effort towards this direction is~\cite{Oliveira07} which investigates
both the completeness and the liveness problems in the measurement of Internet AS graph evolution.

\subsection{Structures beyond the power law}

Degree distribution is a first-order topological property which is based on the connectivity information of
individual nodes. When studying the Internet structure, it is important to look beyond the power
law degree
distribution because networks with exactly the same power
law degree distribution can have completely
different high order properties~\cite{Tangmunarunkit02,Li04,Alderson05,zhou07b}.

High order properties are calculated on  the connectivity information of a pair, a triad or a set of nodes.
High order properties are able to explicitly determine lower order properties whereas the later only
constrain the former. Researchers have introduced many high order topological properties, each of which has a
distinct physical meaning, for example the degree-degree correlation
\cite{Pastor01,newman02,newman03,maslov04} which indicates whether high-degree nodes tend to connect with
high-degree nodes (so-called `assortative mixing') or low-degree nodes (`disassortative mixing'); the
rich-club coefficient \cite{Zhou04a,zhou07b} which quantifies how tightly the best connected nodes connect
with themselves; the clustering coefficient \cite{watts98} which measures the fraction of a node's neighbours
which are neighbours to each other; the average shortest path which is the average hop distance between any
two nodes; the $k$-core decomposition \cite{Carmi07} which reveals a network's underlying hierarchical
structure; and the betweenness which measures how often a node or a link is on the shortest (fewest hop) path between two nodes.

The Internet topology can be describes a jellyfish \cite{tauro01}, where a highly connected core is in the
middle of the cap, and one-degree nodes form its legs. This intuitive model is simple yet very useful as it
concisely illustrates a number of important properties of the Internet, including the dense core (rich-club)
and the large number of low degree nodes (power law) which are directed connected with members of the core
(disassortative mixing). The Internet has a small average distance between any two nodes because the
rich-club functions as a `super' traffic hub which provides a large selection of shortcuts for routing and
the disassortative mixing ensures that the majority of network nodes, which are peripheral low-degree nodes,
are always near the rich-club.

Our knowledge and understanding of the Internet topology have been improved significantly in recent years.
However, it is still profoundly difficult to define the Internet topology and there are many unanswered
questions: What are the key properties that fundamentally characterise the Internet topology?
How do these properties relate to each other? What is the role each property plays on  the network's
function and performance?

It is suggested \cite{Mahadevan06,Mahadevan07} that for the Internet, the second order properties are
sufficient for most practical purposes; while the third order properties essentially reconstruct the Internet
AS and router level topologies exactly. A recent work~\cite{zhou07b} pointed out that for the Internet the
degree distribution and the rich-club coefficient restrict the degree-degree correlation to such a narrow
range, that a reasonable model for the Internet can be produced by considering only the degree distribution
and the rich-club coefficient. Note that although these studies  provide new clues on how to choose
topological properties for consideration in modelling the Internet topology, they do not constitute a
`canonical' set of metrics that are most relevant for the network's function and performance.

\subsection{Modelling Internet topology}

Since the discovery of the power law degree distribution, a  number of models  have been proposed to generate
Internet-like graphs \cite{krapivsky01,albert02,dorogovtsev03,Pastor04,Mitzenmacher05,leskovec07}. Models
from networking community, such as Tier, BRITE \cite{medina00}, GT-ITM (Transit-Stub) and Inet
\cite{winick02}, often  suffer from problems of no (or an
incorrect) power law, inaccurate large-scale
hierarchy, requiring parameter estimation or providing a mechanism
for network evolution; and models from physicists
\cite{Barabasi99,dorogovtsev00,bu02,bianconi03,caldarelli03,krapivsky00} also have problems as they often are
too general and do not incorporate any real network specifics.

In general there are two main approaches for generating topologies of complex networks \cite{rrfs05}. The
equilibrium (top-down) approach is to construct an ensemble of static random graphs reproducing certain
properties of observed networks and then to derive their other properties by the standard methods. The
non-equilibrium approach (bottom-up) tries to mimic the actual dynamics of network growth: if this dynamics
is accurately captured, then the modelling algorithm, when let to run to produce a network of the required
size, will output the topology coinciding with the observations. It is clear that the more ambitious
non-equilibrium approach has the potential to hold the ultimate truth. Classic examples of this approach
include the Barab\'asi-Albert (BA) model \cite{Barabasi99} and the HOT model \cite{Li04}.  Many models owe their origins to the preferential attachment
approach where new links attach to nodes with a probability proportional
to the degree of that node.

The Positive-Feedback Preference (PFP) model proposed in 2004 \cite{Zhou04d} is an example of the
non-equilibrium models for the Internet. The model is an extensive modification of the BA model. It is able
to reproduce a large number of  characteristics (including all topological properties mentioned above) of the
Internet AS topology \cite{krapivsky08,haddadi08, zhou07a}.  It uses two growth mechanisms inspired by
observations on the Internet history data \cite{Pastor01,vazquez02,park04}. Firstly, the model starts from a
small random graph and grows by two coupled actions called \emph{interactive growth}, i.e.~the attachment of
new nodes to old nodes in the existing system and the addition of new links between these old nodes to other
old nodes. This resembles the dynamics that when an Internet service provider (ISP) acquires new customers it
reacts by increasing its number of connections to peering ISPs. Secondly, the preference probability that
node $i$ acquires a new link (from a new node or a peer) is given as a function of the node's degree $k_i$,
\begin{equation} \Pi(i) =
\frac{k_i^{1+\delta\log_{10}{k_i}}}
{\sum_j k_j^{1+\delta\log_{10}{k_j}}}, \delta=0.048.
\label{eq:PFP}
\end{equation}
This is called the \emph{positive-feedback preference}, which means a node's ability of competing for a new
link increases more and more rapidly with its growing number of links, like a positive-feedback loop. The
consequence is that `the rich not just get richer, they get disproportionately richer'. This mechanism
resembles the `winner-takes-all' trend in the Internet development. More recently Chang \emph{et
al}~\cite{chang06} proposed another bottom-up approach for generating Internet AS graph, where the Internet
evolutionary process is modelled by identifying a set of criteria that an AS considers either in establishing
a new peering relationship or in reassessing an existing relationship.

\subsection{Practical responses to Internet power law modelling}

It is suggested \cite{handley06} that the Internet power law structure is relevant to a number of issues, 
such as the severely biased distribution of traffic flow, the slow convergence of BGP routing
tables \cite{labovits01} and the large-scale cascading failure caused by incidents or deliberate
attacks~\cite{park03}. As such, the power law property also provides novel insights into the solutions of
these problems. For example it is shown that the power law property makes it possible to mitigate the
distributed denial of service (DDoS) attacks by implementing route-based filtering on less than 20\% of
AS \cite{park01};  a compact routing scheme based on the power law property requires a
significantly smaller routing table size \cite{krioukov04}, and the power law property is relevant to the
epidemic threshold for a network \cite{wang03}.

Albert et al \cite{Albert2000} have reported that scale-free networks, i.e. networks having power law degree
distributions, are robust to random failures but fragile to targeted attacks. This widely publicised work has
generated a wave of studies on the robustness of various networks. This work, however, has generated some
confusion in the Internet community. It should be noted that the Internet is much different from the generic
BA model used in that study. Firstly the Internet AS topology does not follow a strict power law as in the BA
model and the Internet's high order topological properties are also significantly different from the BA
model. Secondly it is unrealistic, if possible at all, to `attack' an AS node, i.e.~to wipe out an entire AS
network and cut off all its connections with other networks. This is because an AS node can represent a
network of thousands of routers which can spread across a number of continents. And finally it is important
to realise that links on the Internet AS graph can represent different commercial relationships between AS
networks, such that a `path' of adjacent 
links between AS nodes on the Internet AS graph does not
necessarily imply routing `reachability' between the two AS. For example a customer AS does not transit
traffic for its providers.

\subsection{Criticisms and commentary}

The discovery of a power law degree distribution in the Internet topology has attracted a huge amount of
attention and there have been tremendous efforts to measure, characterise and model the Internet topology.
Recent debate suggested that whether the power law degree distribution is an integral property of the
Internet is still an open question. It is vital for researchers to look beyond the power
law property and
appreciate high order properties of the Internet topology.

There are generative models which well reproduce the Internet topology as a pure graph. However the
reachability between two AS nodes on the Internet is not only affected by the underlying connectivity graph,
but also constrained by many other factors, such as routing policies, capacities, demo-geographic
distributions and local structures.  Future Internet models 
should more closely reflect the
Internet specifics in order to produce practically useful results.

As pointed out in \cite{Krioukov07}, there is a need for more interdisciplinary communication among computer scientists, mathematicians, physicists and
engineers. Such communication is much needed to
facilitate the interdisciplinary flow of knowledge and enable the network research community to convert
theoretical results into more practical solutions that matter for real networks, e.g.~performance, revenue
and engineering.

\section{Conclusions}
\label{sec:conclusions}

An obvious question arising from this paper is whether there is a connection
between the power law topology and the power laws observed in traffic levels.
One likely mechanism for such an interaction would come from
considering how traffic
aggregates as a result of the topology.  A starting point might be the work
reported in \cite{arrow04} which combines power law topologies with simulations
involving LRD sources.  Further research in this area might well
be fruitful.

Most authors agree that power law relationships are present in
measurements of network traffic.  Measurements of file size transfers appear
consistent with a heavy-tailed distribution.  Measurements of traffic
levels per unit time and packet inter-arrival times fit the hypothesis
of LRD.  However, this only describes the long time-scale behaviour (at
least below the time-scale where day-to-day non-stationarity affects
measurement).
The behaviour of network traffic at shorter time scales is still
an open question with different authors proposing different models.
On the origin of long-range dependence, consensus appears to have
formed that heavy-tailed distribution of file sizes is the major
cause  but with alterations to the short term behaviour arising from
TCP protocol interactions.  However, some authors give other explanations
for the short term behaviour and
the matter cannot yet be said to be definitively settled.

While many models have been proposed which generate traffic with the appropriate
power law behaviour, it remains to be shown which of these, if any, best fits real
traffic traces. In particular, if LRD is of relevance for  queuing and buffer
behaviour, it is key that the model selected replicates the queuing performance of
the real traffic and this is an important shortcoming. The models proposed to
describe queuing behaviour with long-range dependent input traffic suggest that the
tail of the queue occupancy distribution decays slower than exponentially.
If the study of power laws is to result in a positive effect on
network traffic engineering then: 1) it is important to find a power law based
traffic generation model
which replicates the queuing performance of the real traffic.
2) progress needs to be made in ways to either mitigate the effects of
LRD or to plan a network by allowing for it.

Researchers have also made progress on measuring and modelling the Internet topology at the AS-level. More
complete and accurate measurement data are needed to justify whether the power law degree distribution is
indeed an integral property of the Internet. Much more research work is needed, for example, to
identify the key topological properties that fundamentally characterise the Internet structure and to include
the Internet specifics in topology models. It is encouraging that the power law modelling of Internet
topology have begun to stimulate research which takes advantage of this network structure. There is an
increasing recognition that effective engineering of the global Internet should be based on a detailed
understanding of issues such as the large-scale structure of its underlying physical topology, the manner in
which it evolves over time, and the way in which its constituent components contribute to its overall
function \cite{floyd03}.

In summary, for the research in power laws to truly have an engineering impact on the Internet, reliable and
calibrated models are needed which match the characteristics of real data.   It could be argued that there
has been a certain level of success for topology generation but certainly not for traffic generation. The
models should be capable of application as a design tool to allow engineers to improve real life network
performance.  As yet, this stage of research appears elusive in both fields.

\bibliographystyle{elsarticle-num}
\bibliography{networking}

\begin{thebibliography}{100}
\expandafter\ifx\csname url\endcsname\relax
  \def\url#1{\texttt{#1}}\fi
\expandafter\ifx\csname urlprefix\endcsname\relax\def\urlprefix{URL }\fi
\expandafter\ifx\csname href\endcsname\relax
  \def\href#1#2{#2} \def\path#1{#1}\fi

\bibitem{mitzenmacher04}
M.~Mitzenmacher, A brief history of generative models for power law and
  lognormal distributions, Internet Mathematics 1~(2) (2004) 226--251.

\bibitem{newton05}
M.~Newman, Power laws, {P}areto distributions and {Z}ipf's law, Contemporary
  Physics 46 (2005) 323--351.

\bibitem{leland1993}
W.~E. Leland, M.~S. Taqqu, W.~Willinger, D.~V. Wilson, On the self-similar
  nature of {Ethernet} traffic, in: D.~P. Sidhu (Ed.), Proc. {ACM} {SIGCOMM},
  San Francisco, California, 1993, pp. 183--193.

\bibitem{ltww94}
W.~E. Leland, M.~S. Taqqu, W.~Willinger, D.~V. Wilson, On the self-similar
  nature of {E}thernet traffic (extended version), IEEE/ACM Trans. on
  Networking 2~(1) (1994) 1--15.

\bibitem{Faloutsos99}
M.~Faloutsos, P.~Faloutsos, C.~Faloutsos, On power--law relationships of the
  {I}nternet topology, Computer Comm. Rev. 29 (1999) 251--262.

\bibitem{strogatz01}
S.~H. Strogatz, Exploring complex networks, Nature (London) 410 (2001) 268.

\bibitem{krapivsky01}
P.~L. Krapivsky, S.~Redner, Organization of growing random networks, Phys. Rev.
  E 63 (2001) 066123.

\bibitem{albert02}
R.~Albert, A.~L. Barab\'asi, Statistical mechanics of complex networks, Rev.
  Mod. Phys. 74 (2002) 47--97.

\bibitem{bornholdt02}
S.~Bornholdt, H.~G. Schuster, Handbook of graphs and networks - from the genome
  to the {I}nternet, Wiley-VCH, Weinheim Germany, 2002.

\bibitem{subramanian02}
L.~Subramanian, S.~Agarwal, J.~Rexford, R.~H. Katz, Characterizing the
  {I}nternet hierarchy from multiple vantage points, in: Proc. {IEEE}
  {INFOCOM}, 2002, pp. 618--627.

\bibitem{Chen02}
Q.~Chen, H.~Chang, R.~Govindan, S.~Jamin, S.~Shenker, W.~Willinger, The origin
  of power laws in {I}nternet topologies revisited, in: Proc. {IEEE} {INFOCOM},
  2002, pp. 608--617.

\bibitem{dorogovtsev03a}
S.~N. Dorogovtsev, J.~F.~F. Mendes, Evolution of networks -- from biological
  nets to the {I}nternet and {WWW}, Oxford University Press, Oxford, 2003.

\bibitem{caldarelli03}
G.~Caldarelli, P.~D.~L. Rios, L.~Pietronero, Generalized network growth: from
  microscopic strategies to the real {I}nternet properties,
  arXiv:cond-mat/0307610 v1 (2004).

\bibitem{Pastor04}
R.~Pastor-Satorras, A.~Vespignani, Evolution and structure of the {I}nternet -
  a statistical physics approach, Cambridge University Press, Cambridge, 2004.

\bibitem{chang06}
H.~Chang, S.~Jamin, W.~Willinger, To peer or not to peer: Modeling the
  evolution of the {I}nternet's {AS}-level topology, in: Proc. {IEEE}
  {INFOCOM}, 2006, pp. 1--12.

\bibitem{cleveland2000}
W.~S. Cleveland, D.~Lin, D.~X. Sun, {IP} packet generation: statistical models
  for {TCP} start times based on connection-rate superposition, in: Proc.
  SIGMETRICS, 2000, pp. 166--177.

\bibitem{cao2001}
J.~Cao, W.~S. Cleveland, D.~Lin, D.~X. Sun, On the nonstationarity of
  {I}nternet traffic, SIGMETRICS Perform. Eval. Rev. 29 (2001) 102--112.

\bibitem{b94}
J.~Beran, Statistics for long-memory processes, Chapman \& Hall, {NY}, 1994.

\bibitem{ganesh2004}
A.~Ganesh, N.~O'Connell, D.~Wischik, Big queues, Vol. 1838 of Lecture Notes in
  Mathematics, Springer, 2004.

\bibitem{kriesten1999}
R.~Kriesten, U.~Kaage, F.~Jondral, A unifying view to fractional modeling, in:
  Proc. IEEE GLOBECOM, 1999, pp. 221--226.

\bibitem{pw00}
K.~Park, W.~Willinger, Self-similar network traffic and performance evaluation,
  Wiley, NY, 2000.

\bibitem{gubner2005}
J.~A. Gubner, Theorems and fallacies in the theory of long-range dependent
  processes, IEEE Trans. on Information Theory 51~(3) (2005) 1234--1239.

\bibitem{heath1998}
D.~Heath, S.~Resnick, G.~Samorodnitsky, Heavy tails and long-range dependence
  in on/off processes and associated fluid models, Math. of Oper. Res. 23~(1)
  (1998) 145--165.

\bibitem{cb95}
M.~Crovella, A.~Bestavros, Explaining world wide web traffic self-similarity,
  Tech. rep., TR-95-015 Computer Science Dept., Boston University (1995).

\bibitem{wtsw97}
W.~Willinger, M.~Taqqu, R.~Sherman, D.~Wilson, Self-similarity through high
  variability: statistical analysis of {Ethernet} {LAN} traffic at the source
  level, {IEEE/ACM} Transactions on Networking 5~(1) (1997) 71--86.

\bibitem{liu2005}
Z.~Liu, Long range dependence and heavy tail distributions, Performance
  Evaluation 61~(2--3) (2005) 91--93.

\bibitem{neidhardt98concept}
A.~L. Neidhardt, J.~L. Wang, The concept of relevant time scales and its
  application to queuing analysisof self-similar traffic (or is {H}urst naughty
  or nice?), SIGMETRICS Perform. Eval. Rev. 26~(1) (1998) 222--232.

\bibitem{kara04tenyears}
T.~Karagiannis, M.~Molle, M.~Faloutsos, Long-range dependence: Ten years of
  {I}nternet traffic modeling, IEEE Internet Computing (2004) 57--64.

\bibitem{taqqu1997}
M.~S. Taqqu, V.~Teverovsky, Robustness of {W}hittle type estimators for time
  series with long-range dependence, Stochastic Models 13 (1997) 723--757.

\bibitem{bardet2003}
J.-M. Bardet, G.~Lang, G.~Oppenheim, A.~Phillipe, S.~S. andM. S.~Taqqu,
  Semi-parametric estimation of the long-range dependence parameter: A survey,
  in: P.~Doukhan, G.~Oppenheim, M.~S. Taqqu (Eds.), Theory and applications of
  long-range dependence, Birkh\"{a}user, 2003, pp. 557--577.

\bibitem{pf95}
V.~Paxson, S.~Floyd, Wide-area traffic: the failure of {P}oisson modeling,
  IEEE/ACM Trans. on Networking 3 (1995) 226--244.

\bibitem{cb97}
M.~Crovella, A.~Bestavros, Self-similarity in {WWW} traffic: evidence and
  possible causes, IEEE/ACM Trans. on Networking 5 (1997) 835--846.

\bibitem{fgw98}
A.~Feldmann, A.~C. Gilbert, W.~Willinger, Data networks as cascades:
  investigating the multi-fractal nature of {I}nternet {WAN} traffic, in: Proc.
  {ACM} {SIGCOMM}, 1998, pp. 42--55.

\bibitem{pkc96}
K.~Park, G.~Kim, M.~Crovella, On the relationship between file sizes, transport
  protocols, and self-similar network traffic, Tech. Rep. 1996-016, Boston
  University (1996).

\bibitem{hernandez04}
F.~Hern\'{a}ndez-Campos, J.~S. Marron, G.~Samorodnitsky, F.~D. Smith, Variable
  heavy tails in {I}nternet traffic, Performamce Evaluation 58~(2-3) (2004)
  261--261.

\bibitem{xia2005}
C.~H. Xia, Z.~Liu, M.~S. Squillante, L.~Zhang, N.~Malouch, Web traffic modeling
  at finer time scales and performance implications, Performance Evaluation
  61~(2--3) (2005) 181--201.

\bibitem{park05changetraf}
C.~Park, F.~Hern{\'a}ndez-Campos, J.~S. Marron, F.~D. Smith, Long-range
  dependence in a changing {I}nternet traffic mix, Comput. Networks (2005) 401
  -- 422.

\bibitem{riedi2003}
R.~H. Riedi, Multifractal processes, in: P.~Doukhan, G.~Oppenheim, M.~S. Taqqu
  (Eds.), Theory and applications of long-range dependence, Birkh\"{a}user,
  2003, pp. 625--716.

\bibitem{rl97}
R.~H. Reidi, J.~L. V{\'e}hel, Multifractal properties of {TCP} traffic: a
  numerical study, Tech. Rep. 3129, INRIA, available online at \\ {\tt
  www.stat.rice.edu/$\sim$riedi/Publ/PDF/t.pdf} (1997).

\bibitem{vehel97}
J.~L. V\'{e}hel, R.~Riedi, Fractional Brownian motion and data traffic
  modeling: The other end of the spectrum, Springer, 1997.

\bibitem{erramilli00performance}
A.~Erramilli, O.~Narayan, A.~L. Neidhardt, I.~Saniee, Performance impacts of
  multi-scaling in wide-area {TCP}/{IP} traffic, in: Proc. {IEEE} {INFOCOM},
  2000, pp. 352--359.

\bibitem{zhang03smalltime}
Z.~Zhang, V.~Ribeiro, S.~Moon, C.~Diot, Small-time scaling behaviors of
  {I}nternet backbone traffic: an empirical study, in: Proc. {IEEE} {INFOCOM},
  2003, pp. 1826--1836.

\bibitem{jiang04timeorigin}
H.~Jiang, C.~Dovrolis, The origin of {TCP} traffic burstiness in some time
  scales, Tech. rep., Georgia Institute of Technology, GIT-CERCS-04-09 (2004).

\bibitem{uhlig2004}
S.~Uhlig, High-order scaling and non-stationarity in {TCP} flow arrivals: a
  methodological analysis, Computer Comm. Rev. 34~(2) (2004) 9--24.

\bibitem{hohn2005}
N.~Hohn, D.~Veitch, P.~Abry, Multifractality in {TCP/IP Traffic}: the case
  against, Computer Network Journal 48 (2005) 293--313.

\bibitem{fghw99}
A.~Feldmann, A.~C. Gilbert, P.~Huang, W.~Willinger, Dynamics of {IP} traffic: a
  study of the role of variability and the impactof control, in: Proc. {ACM}
  {SIGCOMM}, 1999, pp. 301--313.

\bibitem{vb00}
A.~Veres, M.~Boda, The chaotic nature of {TCP} congestion control, in: Proc.
  {IEEE} {INFOCOM}, 2000, pp. 1715--1723.

\bibitem{figuerdo2005}
D.~R. Figueiredo, B.~Liu, A.~Feldmann, V.~Misra, D.~Towsley, W.~Willinger, On
  {TCP} and self-similar traffic, Performance Evaluation 61~(2--3) (2005) 129
  -- 141.

\bibitem{hohn2002}
N.~Hohn, D.~Veitch, P.~Abry, Does fractal scaling at the {IP} level depend on
  {TCP} flow arrival processes?, in: Proc. of The Second {I}nternet Measurement
  Workshop, 2002, pp. 63--68.

\bibitem{figueiredo2000auto}
D.~R. Figueiredo, B.~Liu, V.~Misra, D.~Towsley, On the autocorrelation
  structure of {TCP} traffic, Computer Networks 40~(3) (2002) 339--361.

\bibitem{pe97}
J.~M. Peha, Retransmission mechanisms and self-similar traffic models, in:
  {IEEE/ACM/SCS} Comm. Networks and Distributed Systems Modelling and
  Simulation Conference, 1997, pp. 47--52.

\bibitem{veres03tcps}
A.~Veres, Z.~Kenesi, S.~Molnar, G.~Vattay, {TCP}'s role in the propagation of
  self-similarity in the {I}nternet, Computer Communications 26~(8) (2003)
  899--913.

\bibitem{borella98measurement}
M.~Borella, G.~Brewster, Measurement and analysis of long-range dependent
  behavior of {I}nternet packet delay, in: Proc. {IEEE} {INFOCOM}, 1998, pp.
  497--504.

\bibitem{S&V}
R.~V. Sol{\'e}, S.~Valverde, Information transfer and phase transitions in a
  model of {I}nternet traffic, Physica A 289 (2001) 595--605.

\bibitem{arrowsmith2002}
M.~Woolf, D.~K. Arrowsmith, R.~J. Mondragon, J.~M. Pitts, Optimization and
  phase transitions in a chaotic model of data traffic, Phys. Rev. E 66~(4)
  (2002) 046106.

\bibitem{rm99}
R.~Macfadyen, Network traffic characterisation, Master's thesis, UCL (1999).

\bibitem{gw94}
M.~Garrett, W.~Willinger, Analysis, modelling, and generation of self-similar
  {VBR} video traffic, in: Proc. {ACM} {SIGCOMM}, 1994, pp. 269--280.

\bibitem{azn95}
R.~Addie, M.~Zukerman, T.~Neame, Fractal traffic: measurements, modelling, and
  performance evaluation, in: Proc. {IEEE} {INFOCOM}, 1995, pp. 977--984.

\bibitem{n94}
I.~Norros, A storage model with self-similar input, Queuing Systems 16 (1994)
  387--396.

\bibitem{pm96}
M.~Parulekar, A.~Makowski, Tail probabilities for a multiplexer with
  self-similar traffic, in: Proc. {IEEE} {INFOCOM}, 1996, pp. 1452--1459.

\bibitem{tg98}
B.~Tsybakov, N.~Georganas, Self-similar traffic and upper bounds to buffer
  overflow in an {ATM} queue, Performance Evaluation 36 (1999) 359--386.

\bibitem{kherania2005}
A.~A. Kherania, A.~Kumar, Long range dependence in network traffic and the
  closed loop behaviour of buffers under adaptive window control, Performance
  Evaluation 61~(2--3) (2005) 95--127.

\bibitem{chen95model}
Y.~Chen, Z.~Deng, C.~L. Williamson, A model for self-similar {E}thernet {LAN}
  traffic: Design, implementation,and performance implications, in: Summer
  Computer Simulation Conference, 1995, pp. 831--837.

\bibitem{gb96}
M.~Grossglauser, J.~Bolot, On the relevance of long-range dependence in network
  traffic, in: Proc. {ACM} {SIGCOMM}, 1996, pp. 15--24.

\bibitem{re96}
B.~K. Ryu, A.~Elwalid, The importance of long-range dependence of {VBR} video
  traffic in {ATM} traffic engineering: Myths and realities, Computer Comm.
  Rev. 26 (1996) 3--14.

\bibitem{paxson1997}
V.~Paxson, Fast, approximate synthesis of fractional gaussian noise for
  generating self-similar network traffic, Computer Comm. Rev. 27 (1997) 5--18.

\bibitem{roberts96}
J.~Roberts, U.~Mocci, J.~Virtano (Eds.), Broadband network teletraffic.
  Performance evaluation and design of broadband multiservice networks. Final
  report of action {COST} 242, Vol. 1155 of Lecture Notes in Computer Science,
  Springer, 1996.

\bibitem{erramilli94chaotic}
A.~Erramilli, R.~P. Singh, P.~Pruthi, Chaotic maps as models of packet traffic,
  in: Proc. 14th Int. Teletraffic Cong., Vol.~1, 1994, pp. 329--338.

\bibitem{pruthi95}
P.~Pruthi, A.~Erramilli, Heavy-tailed on/off source behaviour and self-similar
  traffic, in: ICC, 1995, pp. 445--450.

\bibitem{burrus98}
C.~Burrus, R.~Gopinath, H.~Guo, Introduction to wavelets and wavelet
  transforms: A primer, Prentice Hall, 1998.

\bibitem{riedi1999}
R.~H. Riedi, M.~S. Crouse, V.~J. Ribeiro, R.~G. Baraniuk, A multifractal
  wavelet model with application to network traffic, {IEEE} Special Issue On
  Information Theory 45(April) (1999) 992--1018.

\bibitem{huebner1998}
F.~Huebner, D.~Liu, J.~M. Fernandez, Queueing performance comparison of traffic
  models for {I}nternet traffic, in: Proc. IEEE GLOBECOM, 1998, pp. 471--476.

\bibitem{clegg2007}
R.~G. Clegg, Simulating {I}nternet traffic with {M}arkov-modulated processes,
  in: Proc. of UK Perf. Eng. Workshop, 2007, pp. 26--37.

\bibitem{quoitin03}
B.~Quoitin, C.~Pelsser, L.~Swinnen, Interdomain traffic engineering with {BGP},
  IEEE Communications Magazine 41~(5) (2003) 122--128.

\bibitem{ripe}
{RIPE}, Routing information service, {RIPE} network coordination center, \quad
  {\tt http://www.ripe.net/}.

\bibitem{nlanr}
{The National Laboratory for Applied Network Research (NLANR)},
  \url{http://moat.nlanr.net/}.

\bibitem{oregon}
{The Route Views project}, {University of Oregon, Eugene},
  \url{http://www.routeviews.org/}.

\bibitem{michigan}
The topology project, {U}niversity of {M}ichigan, {A}nn {A}rbor,
  \url{http://topology.eecs.umich.edu/}.

\bibitem{chang04}
H.~Chang, R.~Govindan, S.~Jamin, S.~Shenker, W.~Willinger, Towards capturing
  representative {AS}-level {I}nternet topologies, Computer Networks Journal
  44~(6) (2004) 737--755.

\bibitem{he07}
Y.~He, G.~Siganos, M.~Faloutsos, S.~Krishnamurthy, A systematic framework for
  unearthing the missing links: Measurements and impact, in: Proc. of 4th
  USENIX Symposium on Networked Systems Design \& Implementation, 2007, pp.
  187--200.

\bibitem{CAIDA}
{The Cooperative Association for {I}nternet Data Analysis (CAIDA)},
  \url{http://www.caida.org/}.

\bibitem{murray01}
M.~Murray, kc~claffy, Measuring the immeasurable: global {I}nternet measurement
  infrastructure, in: Proc. of PAM, 2001, pp. 159--167.

\bibitem{DIMES}
{The DIMES project}, \url{http://www.netdimes.org/}.

\bibitem{hyun03}
Y.~Hyun, A.~Broido, k.~claffy, Traceroute and {BGP} {AS} path incongruities,
  \url{http://www.caida.org/outreach/papers/2003/ASP/}.

\bibitem{Oliveira07}
R.~V. Oliveira, B.~Zhang, L.~Zhang, Observing the evolution of {I}nternet {AS}
  topology, in: Proc. {ACM} {SIGCOMM}, 2007, pp. 313--324.

\bibitem{mahadevan05b}
P.~Mahadevan, D.~Krioukov, M.~Fomenkov, B.~Huffaker, X.~Dimitropoulos,
  K.~Claffy, A.~Vahdat, The {I}nternet {AS}-level topology: Three data sources
  and one definitive metric, Computer Comm. Rev. 36~(1) (2006) 17--26.

\bibitem{cohen06}
R.~Cohen, D.~Raz, The {I}nternet dark matter -- on the missing links in the
  {AS} connectivity map, in: Proc. {IEEE} {INFOCOM}, 2006, pp. 1--12.

\bibitem{Lakhina03}
A.~Lakhina, J.~W. Byers, M.~M.~Crovella, P.~Xie, Sampling biases in {IP}
  topology measurements, in: Proc. {IEEE} {INFOCOM}, Vol.~1, 2003, pp.
  332--341.

\bibitem{Clauset05}
A.~Clauset, C.~Moore, Accuracy and scaling phenomena in {I}nternet mapping,
  Phys. Rev. Lett. 94 (2005) 018701.

\bibitem{DallAsta06}
L.~{Dall-Asta}, I.~{Alvarez-Hamelin}, A.~Barrat, A.~Vazquez, A.~Vespignani,
  Exploring networks with traceroute-like probes: Theory and simulations,
  Theoretical Computer Science 355 (2006) 6--24.

\bibitem{Tangmunarunkit02}
H.~Tangmunarunkit, R.~Govindan, S.~Jamin, S.~Shenker, W.~Willinger, Network
  topology generators: Degree-based vs. structural, in: Proc. {ACM} {SIGCOMM},
  2002, pp. 147--159.

\bibitem{Li04}
L.~Li, D.~Alderson, W.~Willinger, J.~Doyle, A first-principles approach to
  understanding the {I}nternet's router-level topology, in: Proc. {ACM}
  {SIGCOMM}, 2004, pp. 3--14.

\bibitem{Alderson05}
D.~Alderson, L.~Li, W.~Willinger, J.~C. Doyle, Understanding {I}nternet
  topology: Principles, models, and validation, IEEE/ACM Transactions on
  Networking 13~(6) (2005) 1205--1218.

\bibitem{zhou07b}
S.~Zhou, R.~Mondrag\'on, Structural constraints in complex networks, New
  Journal of Physics 9~(173) (2007) 1--11.

\bibitem{Pastor01}
R.~Pastor-Satorras, A.~V\'azquez, A.~Vespignani, Dynamical and correlation
  properties of the {I}nternet, Phys. Rev. Lett. 87~(258701).

\bibitem{newman02}
M.~E.~J. Newman, Assortative mixing in networks, Phys. Rev. Lett. 89 (2002)
  208701.

\bibitem{newman03}
M.~E.~J. Newman, Mixing patterns in networks, Phys. Rev. E 67 (2003) 026126.

\bibitem{maslov04}
S.~Maslov, K.~Sneppenb, A.~Zaliznyaka, Detection of topological patterns in
  complex networks: correlation profile of the {I}nternet, Physica A 333 (2004)
  529--540.

\bibitem{Zhou04a}
S.~Zhou, R.~J. Mondrag\'on, The rich-club phenomenon in the {I}nternet
  topology, IEEE Comm. Lett. 8~(3) (2004) 180--182.

\bibitem{watts98}
D.~J. Watts, S.~H. Strogatz, Collective dynamics of `small-world' networks,
  Nature 393 (1998) 440.

\bibitem{Carmi07}
S.~Carmi, S.~Havlin, S.~Kirkpatrick, Y.~Shavitt, E.~Shir, A model of {I}nternet
  topology using k-shell decomposition, PNAS 104~(27) (2007) 11150--11154.

\bibitem{tauro01}
S.~L. Tauro, C.~Palmer, G.~Siganos, M.~Faloutsos, A simple conceptual model for
  the {I}nternet topology, in: Prof. of Global Internet, 2001, pp. 1667--1671.

\bibitem{Mahadevan06}
P.~Mahadevan, D.~Krioukov, K.~Fall, A.~Vahdat, Systematic topology analysis and
  generation using degree correlations, in: Proc. {ACM} {SIGCOMM}, 2006, pp.
  135--146.

\bibitem{Mahadevan07}
P.~Mahadevan, C.~Hubble, D.~Krioukov, B.~Huffaker, A.~Vahdat, {ORBIS}:
  Rescaling degree correlations to generate annotated {I}nternet topologies,
  in: Proc. {ACM} {SIGCOMM}, 2007, pp. 325--336.

\bibitem{dorogovtsev03}
S.~N. Dorogovtsev, Networks with given correlations,
  \url{http://arxiv.org/abs/cond-mat/0308336v1}.

\bibitem{Mitzenmacher05}
M.~Mitzenmacher, The future of power law research, Internet Mathematics 2~(4)
  (2005) 525--534.

\bibitem{leskovec07}
J.~Leskovec, J.~Kleinberg, C.~Faloutsos, Graph evolution: Densification and
  shrinking diameters, ACM Transactions on Knowledge Discovery from Data 1~(1)
  (2007) 1--40.

\bibitem{medina00}
A.~Medina, I.~Matta, Brite: A flexible generator of {I}nternet topologies,
  Tech. Rep. BU-CS-TR-2000-005, Boston University (2000).

\bibitem{winick02}
J.~Winick, S.~Jamin, Inet-3.0 {I}nternet topology generator, Tech. Rep.
  UM-CSE-TR-456-02, University of Michigan (2002).

\bibitem{Barabasi99}
A.~Barab\'asi, R.~Albert, Emergence of scaling in random networks, Science 286
  (1999) 509.

\bibitem{dorogovtsev00}
S.~N. Dorogovtsev, J.~F.~F. Mendes, Scaling behaviour of developing and
  decaying networks, EuroPhys. Lett. 52~(33) (2000) 33.

\bibitem{bu02}
T.~Bu, D.~Towsley, On distinguishing between {I}nternet power law topology
  generators, in: Proc. {IEEE} {INFOCOM}, 2002, p. 638.

\bibitem{bianconi03}
G.~Bianconi, G.~Caldarelli, A.~Capocci, Number of h-cycles in the {I}nternet at
  the autonomous system level, ArXiv:cond-mat/0310339 (2003).

\bibitem{krapivsky00}
P.~L. Krapivsky, S.~Redner, F.~Leyvraz, Connectivity of growing random
  networks, Phys. Rev. Lett. 85~(4629) (2000) 4629.

\bibitem{rrfs05}
Status report of {IRTF} routing research group's ({RRG}) future domain routing
  ({FDR}) scalability research subgroup ({RR-FS}),
  \url{http://rr-fs.caida.org/} (2005).

\bibitem{Zhou04d}
S.~Zhou, R.~J. Mondrag\'on, Accurately modelling the {I}nternet topology, Phys.
  Rev. E 70 (2004) 066108.

\bibitem{krapivsky08}
P.~Krapivsky, D.~Krioukov, Scale-free networks as pre-asymptotic regimes of
  super-linear preferential attachment, Phys. Rev. E 72 (2008) 026114.

\bibitem{haddadi08}
H.~Haddadi, D.~Fay, A.~Jamakovic, O.~Maennel, A.~W. Moore, R.~Mortier, M.~Rio,
  S.~Uhlig, Beyond node degree: evaluating {AS} topology models, Tech. Rep.
  UCAM-CL-TR-725, University of Cambridge (2008).

\bibitem{zhou07a}
S.~Zhou, G.-Q. Zhang, G.-Q. Zhang, Chinese {I}nternet {AS}-level topology, IET
  Commun. 1~(2) (2007) 209--214.

\bibitem{vazquez02}
A.~V\'azquez, R.~Pastor-Satorras, A.~Vespignani, Large-scale topological and
  dynamical properties of {I}nternet, Phys. Rev. E 65 (2002) 066130.

\bibitem{park04}
S.~T. Park, D.~Pennock, C.~L. Giles, Comparing static and dynamic measurements
  and models of the {I}nternet's {AS} topology, in: Proc. {IEEE} {INFOCOM},
  2004, pp. 1616--1627.

\bibitem{handley06}
M.~Handley, Why the {I}nternet only just works, BT Technology Journal 24~(3)
  (2006) 119--129.

\bibitem{labovits01}
C.~Labovitz, A.~Ahuja, R.~Wattenhofer, S.~Venkatachary, The impact of
  {I}nternet policy and topology on delayed routing convergence, in: Proc.
  {IEEE} {INFOCOM}, 2001, pp. 537--546.

\bibitem{park03}
S.~T. Park, A.~Khrabrov, D.~M. Pennock, S.~Lawrence, C.~L. Giles, L.~H. Ungar,
  Static and dynamic analysis of the {I}nternet's susceptibility to faults and
  attacks, in: Proc. {IEEE} {INFOCOM}, 2003, pp. 2144--2154.

\bibitem{park01}
K.~Park, H.~Lee, On the effectiveness of route-based packet filtering for
  distributed {DoS} attack prevention in power-law {I}nternets, Computer Comm.
  Rev. 31~(4) (2001) 15--26.

\bibitem{krioukov04}
D.~Krioukov, K.~Fall, X.~Yang, Compact routing on {I}nternet-like graphs, in:
  Proc. {ACM} {SIGCOMM}, Vol.~1, 2004, p. 219.

\bibitem{wang03}
Y.~Wang, D.~Chakrabarti, C.~Wang, C.~Faloutsos, Epidemic spreading in real
  networks: an eigenvalue viewpoint, in: Proc. of 22nd International Symposium
  on Reliable Distributed Systems (SRDS), 2003, pp. 25--34.

\bibitem{Albert2000}
R.~Albert, H.~Jeong, A.~Barabasi, Error and attack tolerance of complex
  networks, Nature 406 (2000) 378.

\bibitem{Krioukov07}
D.~Krioukov, F.~Chung, kc~claffy, M.~Fomenkov, A.~Vespignani, W.~Willinger, The
  workshop on {I}nternet topology ({WIT}) report, Computer Comm. Rev. 37~(1)
  (2007) 69--73.

\bibitem{arrow04}
D.~Arrowsmith, M.Woolf, Modelling of {TCP} packet traffic in a large
  interactive growth network, in: Proc. IEEE Systems and Circuits, Vancouver,
  2004, pp. 477--480.

\bibitem{floyd03}
S.~Floyd, E.~Kohler, Internet research needs better models, Computer Comm. Rev.
  33~(1) (2003) 29--34.

\end{thebibliography}
\end{document}